# Half-infinite sampling and its FT

Rui Li

*Abstract*—In the digital world, signals are discrete and finite. The Fourier representation of discrete and finite signals is FT convolution of the finite sampling function and the continuous signal. Conventionally, finite sampling is treated as a segment of infinite sampling. Though this approach perfectly solves the difference between finite and infinite sampling, it has caused much trouble for signal processing. Mathematically, there is a kind of sampling between finite and infinite sampling, and we name this kind of sampling as half-infinite sampling. Theoretically, finite sampling can also be treated as a segment of half-infinite sampling. Because we can derive the Fourier representation of discrete and finite signals from half-infinite sampling, the FTs of several half-infinite samplings are studied. The results show that the FT of half-infinite sampling is more concise than that of infinite sampling. A numerical experiment verified the theoretical derivations successfully, during which step sampling was proposed. Besides, several interesting equations were built.

## I. Introduction

In discrete signal analysis, the spectrum of a discrete and finite signal is the FT convolution of the signal, window function, and sampling function. The FT of the conventional uniform sampling function is the superposition of an infinite number of Dirac $\delta$ functions [1]. According to the theorem of convolution, the FT of a discrete signal is the superposition of an infinite number of spectral lines. In the digital world, the signals processed by computers are finite, and the spectral lines are more or less distorted under the impact of the spectrum leakage effect. Many applications based on FT are affected by spectrum leakage effect, for example, communications, radar positioning, automatic control, pattern recognition, mechanical fault diagnosis, seismic surveying, and so on. The general treatment measure ignores the infinite number of spectrum interference [2]. In most cases, this approach meets the engineering requirements. However, in some cases, likes estimating the frequency of two close-spaced spectral lines [3], the influence of spectrum interference can not be ignored. Therefore, it is of great significance to study the sampling function, especially for accurate calculation.

Shannon's sampling theorem and its extensions are reviewed in reference [4]. The extensions include sampling for more than one variable function, random processes, nonuniform sampling, non-band-limited functions, implicit sampling, generalized functions, sampling with the function, and sampling for general integral transforms [4]. Uniform sampling seems to have been well-studied, whereas nonuniform sampling does not. Nonuniform sampling has been an interesting topic for more than 65 years [5]. Currently, nonuniform sampling has successfully attracted the attention of many physicists in gravitational wave detection. The observation interval deviates from the true periodicity slightly due to clock jitter and time delay. Combined with harsh cosmic environments, nonuniform sampling brings great challenges to weak gravity signal detection. According to the authors' analysis, nonuniform sampling also produces spectrum interference like the spectrum leakage effect. Many techniques were proposed to deal with nonuniform sampling. For example, the spectrum analysis[6], the time-frequency analysis [7], the compressed sensing [8], the nonuniform DFT [9], [10], sampling rate conversion [11], interpolation and reconstruction [12]. The first task of many nonuniform signal processing techniques is to restore uniform signals from the time domain or frequency domain. That is to say, the theory of uniform sampling is the theoretical basis of nonuniform sampling.

Uniform sampling can be divided into finite uniform sampling and infinite uniform sampling. The infinite sampling function is a theoretical function and a mathematical function. In the digital world, signals processed by computers are finite, and finite sampling is used in all aspects of daily life. In reality, finite sampling is treated as infinite sampling multiplied by a window function. Though it perfectly solves the difference between finite and infinite sampling, this approach has caused much trouble for signal processing. For example, the spectrum leakage effect, Gibbs phenomenon [13], [14], and end effect [15]. Mathematically, there is a kind of uniform sampling between finite and infinite sampling, and we name this kind of sampling as half-infinite sampling. Theoretically, finite sampling can also be treated as a segment of half-infinite sampling. Hence, this sampling provides a new way to deal with finite samples, and it has the potential to solve the above problems. The relationship between infinite sampling, finite sampling, and half-infinite sampling is similar to line, line segment, and ray.

The FTs of several half-infinite samplings are derived in this manuscript. Interestingly, they all have only one item, and they are related to the secant function or the cosecant function. An infinite sampling can be regarded as a linear superposition of two half-infinite samplings; hence, we can describe the FT of the infinite uniform sampling with two items. The FT of infinite sampling is the Dirac $\delta$ superposition of infinite items. Hence, the complexity of the FT of the sampling function is

Author: Rui Li; Author's e-mail: rui_li@hust.edu.cn; Author's affiliation: Centre for Gravitational Experiments, School of Physics, MOE Key Laboratory of Fundamental Physical Quantities Measurement and Hubei Key Laboratory of Gravitation and Quantum Physics, PGMF, Huazhong University of Science and Technology, Wuhan 430074, China significantly reduced. Besides, several interesting equations were build based on the re-description of the FT of the sampling function.

## II. BACKGROUND

### A. The correction of symmetric DFT

DFT has two main forms: ordinary and symmetric forms [16]–[18]. The most widely used DFT is ordinary DFT (ODFT), and the well-known FFT [19] is the fast algorithm of ODFT. The ODFT of signal $x(n)$ is defined as:

$$X_o(m) = \sum_{n=0}^{N-1} x(n)e^{-i2\pi mn/N}. \tag{1}$$

Where i is the imaginary unit, $\pi$ is the circumference ratio, e is the Euler's number, $N$ is the signal length, $m$ represents frequency, $n$ represents time. In which $n$ is the index of the sampling function. One important DFT is symmetric DFT (SDFT) [16], also known as unaliased DFT [17] or centered DFT [20]. This form is normally used in interpolation, data compression, and noise removal [18]. The mathematical formula of SDFT is related to the parity of signal length ($N$). When $N$ is odd ($N=2k+1$), the mathematical formula is (2). When $N$ is even ($N=2k$), and the mathematical formula is (3).

$$X_s(m) = \sum_{n=-k}^{k} x(n)e^{-i2\pi mn/N} \tag{2}$$

$$X_s(m) = \sum_{n=-k}^{k-1} x(n)e^{-i2\pi mn/N} \tag{3}$$

When $N$ is odd, the time domain (-$k$, -$k$+1, …, -1, 0, 1, …,$k$-1, $k$) is strictly symmetric to 0. However, when $N$ is even, the time domain (-$k$, -$k$+1, …, -1, 0, 1, …,$k$-2, $k$-1) is not symmetric to 0. Hence, the name of SDFT does not match the fact. One characteristic of even SDFT is $n$ are integers, and it indicates the sampling function of even SDFT is odd sampling (refers to Fig. 1 (a)), and that is the reason for the asymmetry.

In our previous studies [21], we corrected the even SDFT with the even sampling (refers to Fig. 1 (c)). After correction, the time domain of even SDFT is symmetric to 0. One innovation is that $n$ is no longer an integer. The mathematical formula of corrected SDFT is (4).

$$X_s(m) = \sum_{n=-(N-1)/2}^{(N-1)/2} x(n+(N-1)/2)e^{-i2\pi mn/N} \tag{4}$$

### B. The translation property of the sampling function

Assuming the sampling frequency is $f_s$, then the time interval between arbitrary two samples is $\Delta T=1/f_s$. Fig. 1 (a) plots the conventional sampling function (odd sampling function), and the definition of it is

$$s_o(t) = \sum_{n=-\infty}^{\infty} \delta(t-n\Delta T). \tag{5}$$

In which $n$ is an integer, and $\delta$ is the Dirac delta function. The FT of the odd sampling function is

$$S_o(f) = f_s \sum_{j=-\infty}^{\infty} \delta(f-jf_s). \tag{6}$$

The definition of shifted sampling or generalized sampling function is

$$s_r(t) = \sum_{n=-\infty}^{\infty} \delta(t-(n+r)\Delta T). \tag{7}$$

In which $r$ is real-valued. The FT of the shifted sampling can be derived by applying the translation property of FT, as shown in (8).

$$S_r(f) = f_s e^{-i2\pi fr/f_s} \sum_{j=-\infty}^{\infty} \delta(f-jf_s). \tag{8}$$

A special case of shifted sampling is the even sampling function, in which the parameter of $r$ is 0.5. Fig. 1 (c) plots the even sampling function. The FT of this sampling is (9), which can be derived by substituting $r$=0.5 into (8).

$$S_e(f) = S_{r=0.5}(f) = f_s \sum_{j=-\infty}^{\infty} (-1)^j \delta(f-jf_s) \tag{9}$$

Quarterly forward sampling ($r$=0.25) and quarterly backward sampling ($r$=-0.25) are special shifts sampling cases. The FT of quarterly forward and quarterly backward sampling are the following two equations. In which i is the imaginary unit.

$$S_{r=-0.25}(f) = f_s \sum_{j=-\infty}^{\infty} (-i)^j \delta(f-jf_s) \tag{10}$$

$$S_{r=0.25}(f) = f_s \sum_{j=-\infty}^{\infty} (i)^j \delta(f-jf_s) \tag{11}$$

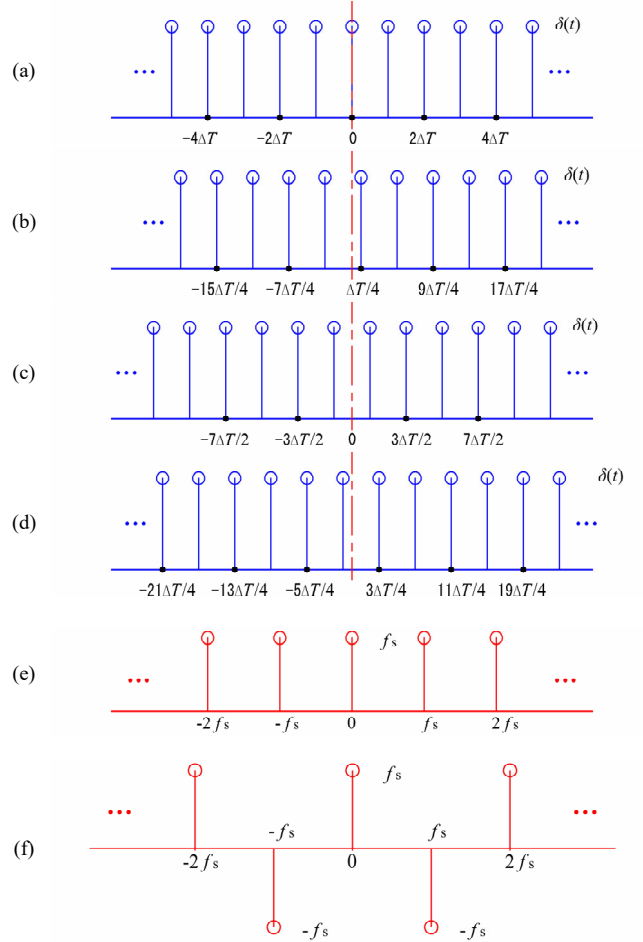

Fig. 1. Different uniform-spaced sampling functions. Subplot (a) is the conventional sampling, also named as the odd sampling; Subplot (b) is the quarterly forward sampling ($r$=0.25); Subplot (c) is the even sampling function ($r$=0.5); Subplot (d) is the quarterly backward sampling ($r$=-0.25); Subplot (e) is the FT of the odd sampling; Subplot (f) is the FT of the even sampling.

## C. The linearity of the sampling function

As introduced in Reference [21], an odd sampling function can be decomposed into an odd sampling function and an even sampling function with the same frequency. That is to say, odd sampling plus even sampling at the same frequency equals frequency-doubled odd sampling, as shown in Fig. 2. According to the linearity of FT, the FT of the frequency-doubled odd sampling is

$$S_o'(f) = S_o(f) + S_e(f) = 2f_s \sum_{j=-\infty}^{\infty} \delta(f - 2jf_s). \quad (12)$$

According to (6), the FT of the frequency-doubled odd sampling is

$$S_o''(f) = 2f_s \sum_{j=-\infty}^{\infty} \delta(f - 2jf_s). \quad (13)$$

The right side of (12) equals the right side of (13). Hence the above derivation is self-contained.

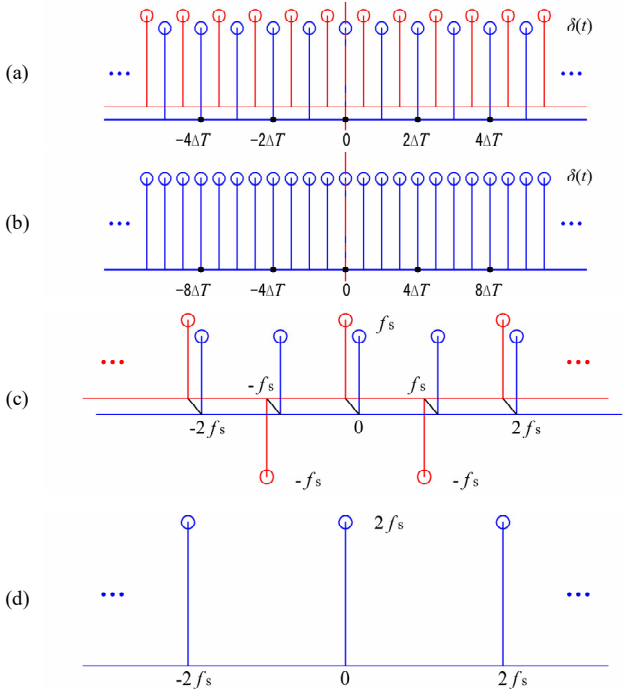

Fig. 2. The linearity of the sampling function. An odd sampling plus an even sampling equals a frequency-doubled odd sampling. Subplot (a) is the progress of the superposition, in which the sampling function in red is the even sampling, and the sampling in blue is the odd sampling; Subplot (b) is the frequency-doubled odd sampling; Subplot (c) is the progress of the frequency superposition; Subplot (d) is the FT of the frequency-doubled odd sampling.

## D. The reversal sampling

The reversal sampling belongs to real-valued uniform sampling, and it can be divided into odd reversal sampling and even reversal sampling, as shown in Fig. 3. The definition formula of the odd reversal sampling is

$$s_{or}(t) = \sum_{n=-\infty}^{\infty} (-1)^n \delta(t - n\Delta T/2). \quad (14)$$

Where the value range of $n$ is $\{n \in \mathbf{N} | -\infty < n < \infty\}$. An odd reversal sampling can be regarded as an odd sampling minus an even sampling. According to the linearity of FT, the FT of odd reversal sampling is

$$S_{or}(f) = 2f_s \sum_{j=-\infty}^{\infty} \delta(f - (2j-1)f_s). \quad (15)$$

The definition formula of the even sampling function is

$$s_{er}(t) = \sum_{n=-\infty}^{\infty} (-1)^n \delta(t - (n+1)\Delta T/2). \quad (16)$$

According to the translation property of FT or the linearity of FT, the FT of even reversal sampling is

$$S_{er}(f) = 2f_s \sum_{j=-\infty}^{\infty} (i)^{(2j-1)} \delta(f - (2j-1)f_s). \quad (17)$$

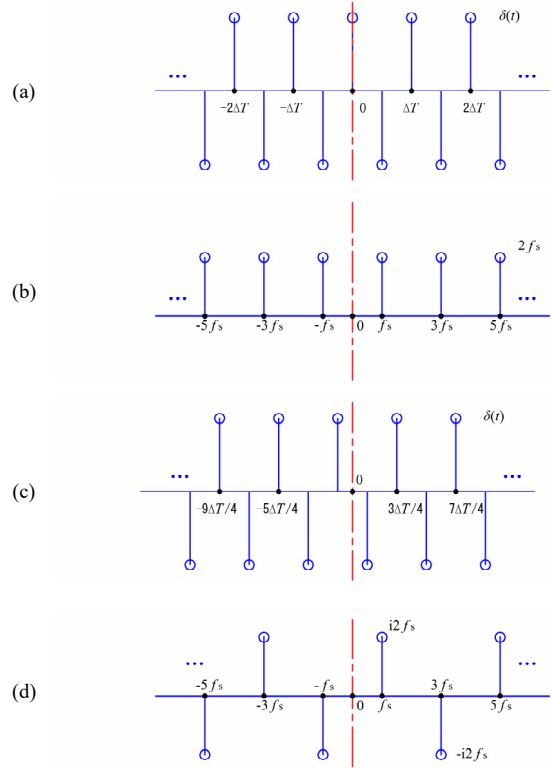

Fig. 3. Two reversal sampling functions. Subplot (a) is the odd reversal sampling; Subplot (b) is the FT of the odd reversal sampling; Subplot (c) is the even reversal sampling; Subplot (d) is the FT of the even reversal sampling.

## III. HALF-INFINITE SAMPLING

### A. Theoretical derivation

In the digital world, finite sampling is treated as infinite sampling multiplied by a window function, and it solves the difference between finite sampling and infinite sampling perfectly. However, this approach has caused much trouble for finite signal processing, for example, the spectrum leakage effect, the Gibbs phenomenon [13], [14], and the end effect [15]. Mathematically, there is a kind of sampling between finite and infinite sampling, and we name this kind of sampling as half-infinite sampling. Theoretically, finite sampling can also be treated as a segment of half-infinite sampling. Hence, this sampling provides a new way to deal with finite samples. The relationship between infinite sampling, finite sampling, and half-infinite sampling is similar to line, line segment, and ray. Six half-infinite samplings are plotted in Fig. 4. The definition formulas of the six samplings are shown in (18).

> < 3

Case 1 $\quad s_{h1}(t) = \sum_{n=0}^{\infty} \delta(t - n\Delta T)$

Case 2 $\quad s_{h2}(t) = \sum_{n=1}^{\infty} \delta(t - n\Delta T)$

Case 3 $\quad s_{h3}(t) = \sum_{n=-\infty}^{-1} \delta(t - n\Delta T)$ (18)

Case 4 $\quad s_{h4}(t) = \sum_{n=-\infty}^{0} \delta(t - n\Delta T)$

Case 5 $\quad s_{h5}(t) = \sum_{n=-\infty}^{0} \delta(t - (n - 0.5)\Delta T)$

Case 6 $\quad s_{h6}(t) = \sum_{n=0}^{\infty} \delta(t - (n + 0.5)\Delta T)$

The six sampling functions listed above have the following four relationships.

$$s_{h1}(t) = s_{h6}(t + \Delta T/2) \quad (19)$$
$$s_{h2}(t) = s_{h6}(t - \Delta T/2) \quad (20)$$
$$s_{h3}(t) = s_{h5}(t + \Delta T/2) \quad (21)$$
$$s_{h4}(t) = s_{h5}(t - \Delta T/2) \quad (22)$$

Assuming the FT of the six samplings are $S_{h1}$, $S_{h2}$, $S_{h3}$, $S_{h4}$, $S_{h5}$, $S_{h6}$, and according to the translation property of FT, we have the following four equations.

$$S_{h1}(f) = e^{i2\pi f \Delta T/2} S_{h6}(f) \quad (23)$$
$$S_{h2}(f) = e^{-i2\pi f \Delta T/2} S_{h6}(f) \quad (24)$$
$$S_{h3}(f) = e^{i2\pi f \Delta T/2} S_{h5}(f) \quad (25)$$
$$S_{h4}(f) = e^{-i2\pi f \Delta T/2} S_{h5}(f) \quad (26)$$

When compared $s_{h1}$ with $s_{h2}$ and $s_{h3}$ with $s_{h4}$, we have the following two equations.

$$s_{h1}(t) - s_{h2}(t) = \delta(t) \quad (27)$$
$$s_{h4}(t) - s_{h3}(t) = \delta(t) \quad (28)$$

According to the linearity of FT, we get:

$$S_{h1}(f) - S_{h2}(f) = 1, \quad (29)$$
$$S_{h4}(f) - S_{h3}(f) = 1. \quad (30)$$

Substituting (23) and (24) into (29), substituting (25) and (26) into (30), we have the following two equations.

$$e^{i\pi f \Delta T} S_{h6}(f) - e^{-i\pi f \Delta T} S_{h6}(f) = 1 \quad (31)$$
$$e^{-i\pi f \Delta T} S_{h5}(f) - e^{i\pi f \Delta T} S_{h5}(f) = 1 \quad (32)$$

After simplification, we get the FT of $s_{h6}$ and $s_{h5}$, as shown in (33) and (34).

$$S_{h6}(f) = \frac{1}{2i \sin(\pi f \Delta T)} \quad (33)$$
$$S_{h5}(f) = \frac{1}{2i \sin(-\pi f \Delta T)} \quad (34)$$

Substituting (33) and (34) into (23)-(26), we get the FT of the other four half-infinite samplings, as shown in (35)-(38).

$$S_{h1}(f) = \frac{e^{i\pi f \Delta T}}{2i \sin(\pi f \Delta T)} \quad (35)$$
$$S_{h2}(f) = \frac{e^{-i\pi f \Delta T}}{2i \sin(\pi f \Delta T)} \quad (36)$$
$$S_{h3}(f) = \frac{e^{i\pi f \Delta T}}{2i \sin(-\pi f \Delta T)} \quad (37)$$
$$S_{h4}(f) = \frac{e^{-i\pi f \Delta T}}{2i \sin(-\pi f \Delta T)} \quad (38)$$

### B. Theoretical validation

The odd sampling function plus the even sampling function at the same frequency equals an odd sampling function whose frequency is doubled. The half-infinite sampling functions discussed above have the same linearity. A series of derivations is done to validate the linearity of half-infinite samplings.

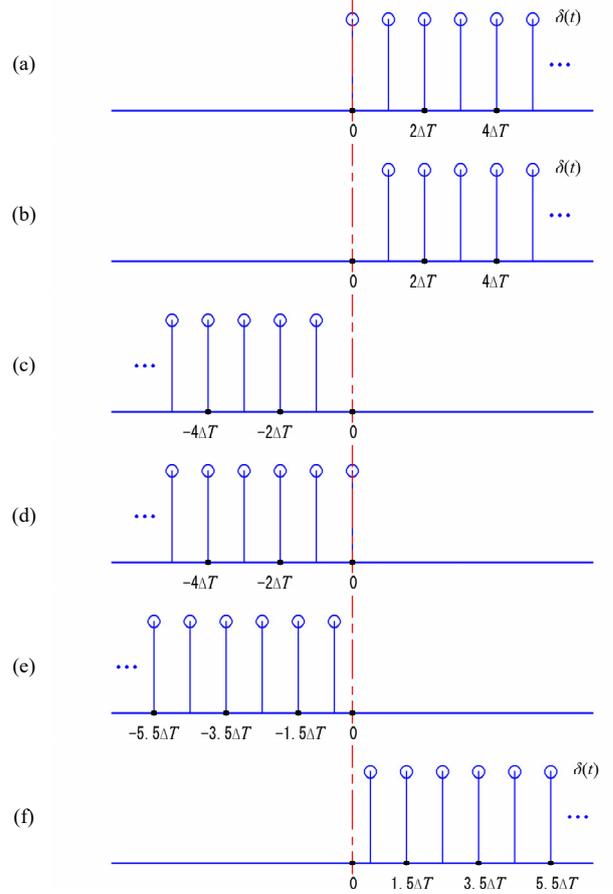

Fig. 4. Various half-infinite sampling functions. In which six half-infinite sampling functions are plotted. Subplot (a) is the case 1; subplot (b) is the case 2; subplot (c) is the case 3; subplot (d) is the case 4; subplot (e) is the case 5; subplot (f) is the case 6.

In the time domain, case 1 plus case 6 equals a frequency-doubled case 1; case 2 plus case 6 equals a frequency-doubled case 2; case 3 plus case 5 equals a frequency-doubled case 3; case 4 plus case 5 equals a frequency-doubled case 4. To validating the frequency domain linearity, two examples of frequency domain superposition are described below.

*1) Case 1 plus case 6*

According to (35), the FT of the frequency-doubled case 1 is

$$S_{h1}'(f) = \frac{e^{i\pi f \Delta T/2}}{2i \sin(\pi f \Delta T/2)} \quad (39)$$

According to the linearity of FT, the FT of the frequency-doubled case 1 is

$$S_{h1}''(f) = S_{h1}(f) + S_{h6}(f) \quad (40)$$

Substituting (33) and (35) into (40), we have:

$$S_{h1}''(f) = \frac{e^{i\pi f \Delta T}}{2i \sin(\pi f \Delta T)} + \frac{1}{2i \sin(\pi f \Delta T)}$$
$$= \frac{e^{i\pi f \Delta T/2}(e^{i\pi f \Delta T/2} + e^{-i\pi f \Delta T/2})}{4i \sin(\pi f \Delta T/2) \cos(\pi f \Delta T/2)} \quad (41)$$

Substituting Euler's equation into (41), we have:

$$S_{h1}''(f) = \frac{e^{i\pi f \Delta T/2}}{2i \sin(\pi f \Delta T/2)} = S_{h1}'(f) \quad (42)$$

*2) Case 2 plus case 6*

According to (36), the FT of the frequency-doubled case 2 is

$$S_{h2}'(f) = \frac{e^{-i\pi f \Delta T/2}}{2i \sin(\pi f \Delta T/2)} \quad (43)$$

According to the linearity of FT, the FT of the frequency-doubled case 2 is

$$S_{h2}''(f) = S_{h2}(f) + S_{h6}(f) \qquad (44)$$

Substituting (33) and (36) into (44), we have:

$$S_{h2}''(f) = \frac{e^{-i\pi f \Delta T}}{2i\sin(\pi f \Delta T)} + \frac{1}{2i\sin(\pi f \Delta T)}$$
$$= \frac{e^{-i\pi f \Delta T/2}(e^{-i\pi f \Delta T/2} + e^{i\pi f \Delta T/2})}{4i\sin(\pi f \Delta T/2)\cos(\pi f \Delta T/2)} \qquad (45)$$

Substituting Euler's equation into (45), we have:

$$S_{h2}''(f) = \frac{e^{-i\pi f \Delta T/2}}{2i\sin(\pi f \Delta T/2)} = S_{h2}'(f) \qquad (46)$$

*C. New equations*

An infinite sampling can be regarded as a linear superposition of two half-infinite samplings, and at the same time, something interesting appears. As shown in Fig. 5, the even sampling can be regarded as case 5 plus case 6. According to the linearity of FT, we obtain the following equation.

$$\frac{1}{2i\sin(\pi f \Delta T)} + \frac{1}{2i\sin(-\pi f \Delta T)} = f_s \sum_{j=-\infty}^{\infty} (-1)^j \delta(f - jf_s) \qquad (47)$$

Where $f_s = 1/\Delta T$. When $k = f \times \Delta T$ is not an integer, the left side of equation (47) equals zero, and the right side of equation (47) also equals zero. Under this condition, the above equation holds. When $k = f \times \Delta T$ is an integer ($k=j$), the two items on left side of (47) have singularity, and the two items can not be added directly; the right side of (47) is proportional to Dirac delta $\delta(0)$, which is not a infinite number, and there is no scientific conclusion. If (47) holds at the singular points, it surpasses the existing knowledge of complex number operations, and the nature behind this equation needs further study. For $k = f \times \Delta T$ is an odd number, we have

$$\frac{1}{2i\sin(k\pi)} + \frac{1}{2i\sin(-k\pi)} = -f_s \delta(f - kf_s). \qquad (48)$$

For $k = f \times \Delta T$ is an even number, we have

$$\frac{1}{2i\sin(k\pi)} + \frac{1}{2i\sin(-k\pi)} = f_s \delta(f - kf_s). \qquad (49)$$

The two equations above are related to the imaginary unit, circumference ratio, sine function, sampling frequency, and Dirac delta function. They are derived from and different from Euler's equation. Currently, there is no convincing evidence that these two equations are right or wrong. Similar to (47), we can get the following four relationships from half-infinite sampling.

$$\frac{e^{i\pi f \Delta T}}{2i\sin(\pi f \Delta T)} + \frac{e^{-i\pi f \Delta T}}{2i\sin(-\pi f \Delta T)} = f_s \sum_{j=-\infty}^{\infty} \delta(f - jf_s) + 1 \qquad (50)$$

$$\frac{e^{-i\pi f \Delta T}}{2i\sin(\pi f \Delta T)} + \frac{e^{i\pi f \Delta T}}{2i\sin(-\pi f \Delta T)} = f_s \sum_{j=-\infty}^{\infty} \delta(f - jf_s) - 1 \qquad (51)$$

$$\frac{e^{i\pi f \Delta T}}{2i\sin(\pi f \Delta T)} + \frac{e^{i\pi f \Delta T}}{2i\sin(-\pi f \Delta T)} = f_s \sum_{j=-\infty}^{\infty} \delta(f - jf_s) \qquad (52)$$

$$\frac{e^{-i\pi f \Delta T}}{2i\sin(\pi f \Delta T)} + \frac{e^{-i\pi f \Delta T}}{2i\sin(-\pi f \Delta T)} = f_s \sum_{j=-\infty}^{\infty} \delta(f - jf_s) \qquad (53)$$

The physical meaning of (52) is that the even sampling function shifts $\Delta T/2$ to the left gets the odd sampling function, and the physical meaning of (53) is that the even sampling function shifts $\Delta T/2$ to the right also gets the odd sampling function. Although the four equations above could be further simplified and something new may emerge, caution should be exercised in the derivation process.

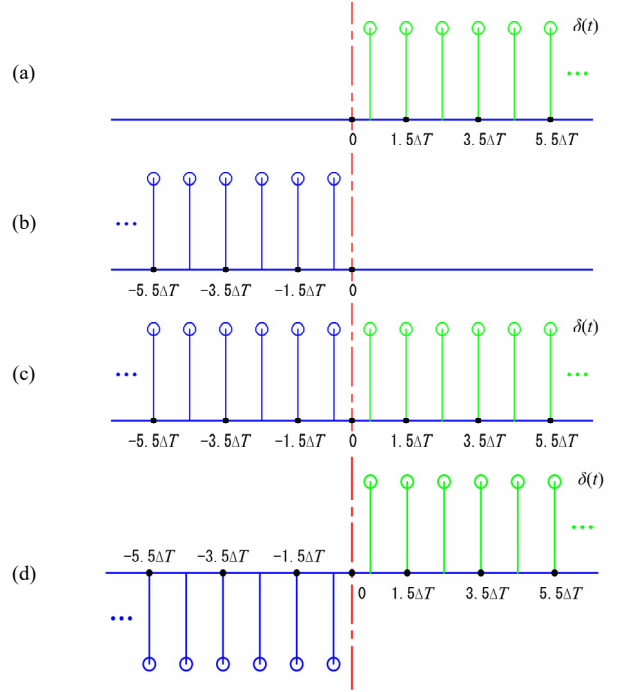

Fig. 5. The linear superposition of two half-infinite samplings. The even sampling can be treated as case 5 plus case 6 half-infinite sampling, and case 6 minus case 5 composes a step sampling. Subplot (a) is the half-infinite sampling of case 6; subplot (b) is the half-infinite sampling of case 5; subplot (c) is the even sampling; subplot (d) is a step sampling.

*D. Numerical validation*

Step sampling is constructed to test the theory of half-infinite sampling indirectly. One advantage of step sampling is easy to implement. The definition formula of step sampling is :

$$s_s(t) = \sum_{n=-\infty}^{\infty} \text{sign}(t) \times \delta(t - (n + 0.5)\Delta T). \qquad (54)$$

Where sign() is the sign function. As shown in Fig. 5, step sampling can be regarded as case 6 minus case 5. According to the linearity of FT, the FT of step sampling is

$$S_s(f) = \frac{-i}{\sin(\pi f \Delta T)}. \qquad (55)$$

$S_s(f)$ is a periodic function whose period is $2f_s$, the same as even sampling. A rectangular window is adopted to test step sampling, assuming $N$ samples ($N$ is even) are obtained. According to the theory of convolution, the FT of the $N$ samples is

$$X(f) = W(f) \otimes S_s(f). \qquad (56)$$

Where $W(f)$ is the FT of the rectangular window, "$\otimes$" represents convolution. The FT of the continuous rectangular window is

$$W(f) = \frac{N}{f_s} \text{sinc}(N\pi f/f_s) = N\Delta T\, \text{sinc}(N\pi f \Delta T). \qquad (57)$$

We can see that $W(f)$ has no periodicity. Substituting (55) and (57) into (56), DTFT of the $N$ samples becomes

$$X(f) = N\Delta T\, \text{sinc}(N\pi f \Delta T) \otimes \frac{-i}{\sin(\pi f \Delta T)}. \qquad (58)$$

According to the theory of convolution, the periodicity of $X(f)$ is the same as that of $S_s(f)$. Another characteristic of $X(f)$ is that it has only imaginary parts.

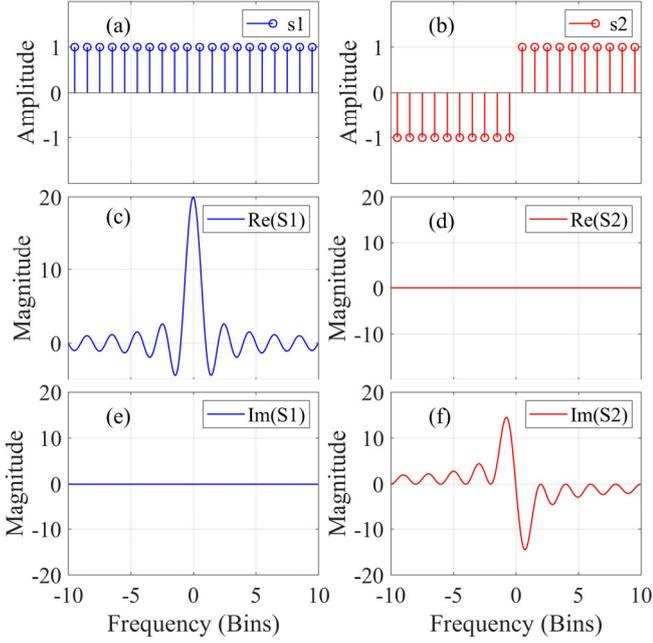

Fig. 6. Symmetric DTFT spectrum comparison of a discrete rectangular window. In which signal length *N* is 20. The signal used is a rectangular window. The sampling function of signal s1 is even sampling, whereas the sampling function of signal s2 is step sampling. S1 is the symmetric discrete-time Fourier transform of s1, and S2 is the symmetric discrete-time Fourier transform of s2. Subplot (a) plots the samples of s1; subplot (b) plots the samples of s2; subplot (c) is the real spectrum of S1; subplot (d) is the real spectrum of S2; subplot (e) is the imaginary spectrum of S1; subplot (f) is the imaginary spectrum of S2.

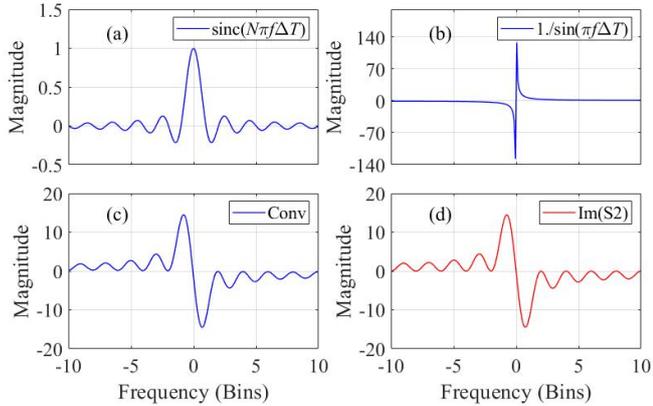

Fig. 7. The numerical test for formula (58). The output result proves our derivation. Subplot (a) is the FT of the continuous rectangular window; subplot (b) is the FT of step sampling; subplot (c) is the FT convolution of the step sampling and rectangular window; subplot (d) is the imaginary spectrum of S2, which is a subplot in Fig. 9. Subplot (c) is the same as subplot (d), and it proves our derivations. In this simulation, the sampling frequency ($f_s$) is 20 Hz, *N* is 20, the value range of frequency is $\{f \in \mathbf{R} | -N/2 < f < N/2\}$, and the frequency step is 0.1 bins.

Symmetric DFT has more FT properties and is more suitable as the discrete form of FT [21]. Hence, symmetric DFT and its zero-padding [22] were applied to approximating the symmetric DTFT spectrum. The symmetric DTFT spectrums of the discrete rectangular window are plotted in Fig. 6, in which signal length (*N*) is 20. The real spectrum is constant 0, and the imaginary spectrum is oddly symmetric. A numerical test is done to prove formula (58). The output result in Fig. 7 shows the convolution of the two functions equals the imaginary spectrum of s2. It successfully proves the FT of the step sampling, as well as the FTs of half-infinite sampling functions.

We can understand the two symmetric DTFT spectrums from the ordinary DTFT spectrum of a discrete rectangular window (signal s3), which is plotted in Fig. 8. The real spectrum of s3 is similar to that of s1, and the imaginary spectrum of s3 is similar to the imaginary spectrum of s2. Hence, the strange curve of step sampling (subplot (f) of Fig. 6) has source. If we add signal s1 to signal s2, we obtain a new rectangular window (s4), as shown in Fig. 8 (b). We can see the real and imaginary spectrum of s4 are regular, whereas, the real and imagniary spectrum of s3 are not. The difference between them can be explained by the FT translation.

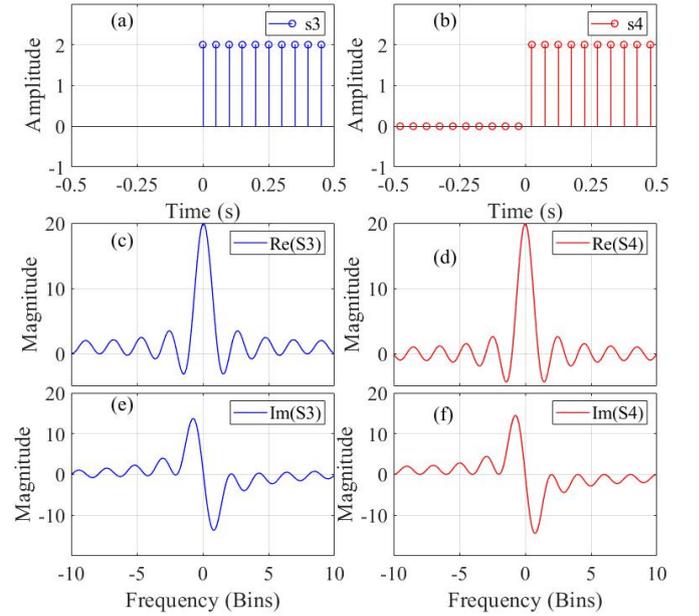

Fig. 8. Two DTFT spectrums comparison. Subplot (a) is the discrete rectangular window (s3) of the ordinary DFT; subplot (b) is the superposition of s1 and s2; subplot (c) is the real spectrum of s3 that calculated with ordinary DTFT; subplot (d) is the real spectrum of s4 that calculated with symmetric DTFT; subplot (e) is the imaginary spectrum of s3 that calculated with ordinary DTFT; subplot (f) is the imaginary spectrum of s4 that calculated with symmetric DTFT. The imaginary spectrum of s4 is similar to that of s3, which proves the strange curve of step sampling (subplot (f) of Fig. 6) has source, and it proves our theory of half-infinite sampling indirectly.

### E. Half-infinite reversal sampling

The concept of half-infinite reversal sampling comes from reversal sampling, which can be regarded as the subtraction of two half-infinite sampling functions. Four half-infinite reversal samplings are plotted in Fig. 9. According to the linearity or translation property, the FTs of the four samplings are the following four formulas.

$$
\begin{aligned}
\text{Case 1} \quad & S_{\text{hr1}}(f) = \frac{e^{i\pi f \Delta T/2}}{2\cos(\pi f \Delta T/2)} \\
\text{Case 2} \quad & S_{\text{hr2}}(f) = \frac{e^{-i\pi f \Delta T/2}}{2\cos(-\pi f \Delta T/2)} \\
\text{Case 3} \quad & S_{\text{hr3}}(f) = \frac{1}{2\cos(\pi f \Delta T/2)} \\
\text{Case 4} \quad & S_{\text{hr4}}(f) = \frac{1}{2\cos(-\pi f \Delta T/2)}
\end{aligned}
\quad (59)
$$

The even reversal sampling can be regarded as case 4 half-infinite reversal sampling minus case 3, as shown in Fig. 9 (e). According to (17), we have the following equation.

$$\frac{1}{2\cos(-\pi f \Delta T)} - \frac{1}{2\cos(\pi f \Delta T)}$$
$$= f_s \sum_{j=-\infty}^{\infty} (i)^{(2j-1)} \delta(f - (j-0.5)f_s). \quad (60)$$

When $2f \times \Delta T \neq 2k+1$, and $\{k \in \mathbf{Z}|-\infty<k<\infty\}$, the left side of (60) is zero, the right side of (60) is also zero. Under this condition, the above equation holds. When $2f \times \Delta T = 4k+1$, and $\{k \in \mathbf{Z}|-\infty<k<\infty\}$, we have

$$\frac{1}{2\cos(-\pi/2 + 2k\pi)} - \frac{1}{2\cos(\pi/2 + 2k\pi)}$$
$$= i f_s \delta(f - (4k+1)f_s/2). \quad (61)$$

When $2f \times \Delta T = 4k+3$, and $\{k \in \mathbf{Z}|-\infty<k<\infty\}$, we have

$$\frac{1}{2\cos(-3\pi/2 + 2k\pi)} - \frac{1}{2\cos(3\pi/2 + 2k\pi)}$$
$$= -i f_s \delta(f - (4k+3)f_s/2). \quad (62)$$

The left side of (61) and (62) are real and infinity; the right side of (61) and (62) are imaginary and infinity. Hence, there is something unknown behind the two equations.

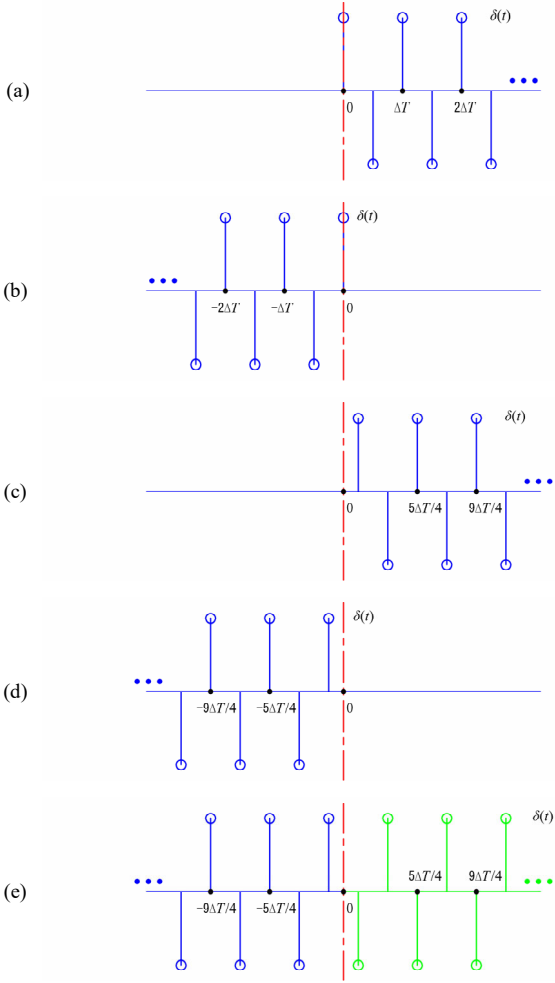

Fig. 9. Various half-infinite reversal sampling functions. Subplot (a) is the case 1 half-infinite reversal sampling; subplot (b) is the case 2 half-infinite reversal sampling; subplot (c) is the case 3 half-infinite reversal sampling; subplot (d) is the case 4 half-infinite reversal sampling; subplot (e) is the even reversal sampling, and it can be regarded as case 4 half-infinite reversal sampling minus case 3 half-infinite reversal sampling.

## IV. DISCUSSION

The Gibbs phenomenon describes a peculiar phenomenon of the Fourier series. The Fourier series of a piecewise continuously differentiable periodic function behaves at a jump discontinuity [13]. Step sampling provides a possible way of solving discontinuity.

## V. REFERENCES

[1] A. V Oppenheim and R. W. Schafer, "Discrete Time Signal Processing 2nd Edition," *Book*. 1998.
[2] D. Kang, X. Ming, and Y. Zhijiang, *The theory and technology of discrete spectrum correction*. Chinese Science Publishing Company, 2008.
[3] J. Luo and M. Xie, "Phase difference methods based on asymmetric windows," *Mechanical Systems and Signal Processing*, 2015, doi: 10.1016/j.ymssp.2014.08.023.
[4] A. J. Jerri, "The Shannon sampling theorem—Its various extensions and applications: A tutorial review," *Proceedings of the IEEE*, vol. 65, no. 11, pp. 1565–1596, 1977, doi: 10.1109/PROC.1977.10771.
[5] J. Yen, "On nonuniform sampling of bandlimited signals," *IRE Trans. Circuit Theory*, vol. CT-3, 1956.
[6] P. Babu and P. Stoica, "Spectral analysis of nonuniformly sampled data – a review," *Digital Signal Processing*, vol. 20, no. 2, pp. 359–378, 2010, doi: https://doi.org/10.1016/j.dsp.2009.06.019.
[7] R. H. Jones, "Fitting a continuous time autoregression to discrete data," *Applied Time Series Analysis II*, pp. 651–682, 1981, doi: https://doi.org/10.1016/B978-0-12-256420-8.50026-5.
[8] Y. C. Eldar, *Sampling theory: Beyond bandlimited systems*. 2014.
[9] S. Bagchi and S. K. Mitra, *The Nonuniform Discrete Fourier Transform and Its Applications in Signal Processing*. Kluwer Academic Publishers, 1999.
[10] F. A. Marvasti, "Nonuniform Sampling: Theory and Practice," *computer*, 2001.
[11] P. Martínez-Nuevo, "Nonuniform Sampling Rate Conversion:An Efficient Approach," *IEEE Transactions on Signal Processing*, vol. 69, pp. 2913–2922, 2021, doi: 10.1109/TSP.2021.3079802.
[12] A. Ignjatović, C. Wijenayake, and G. Keller, "Chromatic Derivatives and Approximations in Practice—Part II: Nonuniform Sampling, Zero-Crossings Reconstruction, and Denoising," *IEEE Transactions on Signal Processing*, vol. 66, no. 6, pp. 1513–1525, 2018, doi: 10.1109/TSP.2017.2787149.
[13] T. A. A. B. and H. S. Carslaw, "Introduction to the Theory of Fourier's Series and Integrals," *The Mathematical Gazette*, 1950, doi: 10.2307/3608695.
[14] A. J. Jerri, *The Gibbs Phenomenon in Fourier Analysis, Splines and Wavelet Approximations*. 1998.
[15] C. S. Turner, "An efficient analytic signal generator," *IEEE Signal Processing Magazine*, 2009, doi: 10.1109/MSP.2009.932794.
[16] R. Strawderman, W. L. Briggs, and V. E. Henson, "The DFT: An Owner's Manual for the Discrete

| | |
|---|---|
| | Fourier Transform," *Journal of the American Statistical Association*, 1995, doi: 10.2307/2669724. |
| [17] | A. Dutt, "Fast Fourier Transforms for Nonequispaced Data," Yale University, 1993. |
| [18] | P. J. Olver, "Topics in Fourier Analysis : DFT & FFT , Wavelets , Laplace Transform," 2018. |
| [19] | J. W. Cooley and J. W. Tukey, "An Algorithm for the Machine Calculation of Complex Fourier Series," *Source: Mathematics of Computation*, 1965, doi: 10.2307/2003354. |
| [20] | D. H. Mugler, "The centered discrete Fourier transform and a parallel implementation of the FFT," 2011, doi: 10.1109/ICASSP.2011.5946834. |
| [21] | R. Li, "A promotion for odd symmetric discrete Fourier transform," 2021. |
| [22] | R. Li, J. Xuan, and T. Shi, "Frequency estimation based on symmetric discrete Fourier transform," *Mechanical Systems and Signal Processing*, vol. 160, p. 107911, 2021, doi: https://doi.org/10.1016/j.ymssp.2021.107911. |